\documentclass[aps,epsfig,preprint,groupedaddress]{revtex4-1}
\usepackage{amssymb}
\usepackage{mathtools}
\usepackage{graphicx}
\usepackage{color}
\usepackage{hyperref}
\usepackage{fullpage}
\usepackage{comment}
\makeatletter
\gdef\@ptsize{0}
\let\@currsize\normalsize

\begin{document}
\title{Effect of time varying transmission rates on the coupled dynamics of epidemic and awareness over a multiplex network}
\author {Vikram Sagar$^{1}$ Yi Zhao$^{1}$}
\email{zhao.yi@hit.edu.cn}
\author{Abhijit Sen$^{2}$}
\affiliation{Harbin Institute of Technology, Shenzhen 518055 ,China$^{1}$ \\ Institute For Plasma Research, Gandhinagar 382428, India$^2$}
\begin{abstract}
A non-linear stochastic model is presented to study the effect of time variation of transmission rates on the co-evolution of epidemics and its corresponding awareness over a two layered multiplex network. In the model, the infection transmission rate of a given node in the epidemic layer depends upon its awareness probability in the awareness layer. Similarly, the infection information transmission rate of a node in the awareness layer depends upon its infection probability in the epidemic layer. The spread of disease resulting from physical contacts is described in terms of a SIS (Susceptible Infected Susceptible) process over the epidemic layer and the spread of information about the disease outbreak is described in terms of an UAU (Unaware Aware Unaware) process over the virtual interaction mediated awareness layer. The time variation of the transmission rates and the resulting co-evolution of these mutually competing processes is studied in terms of a network topology dependent parameter ($\alpha$). Using a second order linear theory it is shown that in the continuous time limit, the co-evolution of these processes can be described in terms of damped and driven harmonic oscillator equations. From the results of a Monte-Carlo simulation, it is shown that for a suitable choice of the parameter$(\alpha)$, the two processes can either exhibit sustained oscillatory or damped dynamics. The damped dynamics corresponds to the endemic state. Further, for the case of an endemic state it is shown that inclusion of the awareness layer significantly lowers the disease transmission rate and reduces the size of the epidemic. The endemic state infection probability of a given node corresponding to the damped dynamics is found to have a dependence upon both the transmission rates as well as on the absolute intra-layer and relative inter-layer degrees of the individual nodes.            
\end{abstract}
\maketitle

\section{Introduction}
\subsection{Leading Paragraph}
{\bf The advent of socializing portals has led to a surge in the virtual contacts mediated interactions among individuals. Such alternate modes of interactions act in conjugation with those arising from persistent real physical contacts. The intricate coupling of these aforementioned interactions can affect the outbreak threshold and even alter the progression of the resultant dynamical processes such as the spread of epidemics and its corresponding awareness. In earlier works, the dynamics of such mutually coupled processes have been studied assuming the respective rates of transmission to be time invariant and mutually exclusive of each other. However, in real life the interaction among such interdependent processes can cause temporal modulation of the respective rates of transmissions depending upon the state of participating individuals. In this work, a non-liner stochastic model is presented to study the dynamics of such interdependent processes over a multiplex network. The model highlights the role of network topology in the self-consistent temporal modulation of respective rates of transmission. The theoretical and Monte Carlo simulation results delineate the interplay of different factors in the outbreak and subsequent spread of such mutually dependent processes.}

The spread of epidemics along with information dissemination are two intriguing and traditionally independent topics of research in the study of complex networks. However, the last few years have witnessed an ever increasing diversification in the communication services due to rapid advances in the field of wireless technology which has interlinked different mobile devices with social networks. Such couplings have resulted in alternate modes of communication which in turn have facilitated the virtual contacts mediated dissemination of information\cite{mobile}. These modes of information dissemination are complementary to the traditional ones like word of mouth which result from physical contacts. The spread of disease information induced awareness over such modes can have profound effects on the threshold of the epidemic outbreak as well as upon its subsequent size. The prominent real life examples highlighting the role of awareness in the spread of epidemics includes the outbreak of severe acute respiratory syndrome (SARS) in the year 2003 in China \cite{Sars}, spread of H1N1 virus and bird flu, etc.. In these scenarios, the dissemination of the epidemic outbreak information results in disease awareness among the individuals, thereby prompting them to take preventive measures. The resultant preventive measures undertaken by the individuals may include wearing masks, improving personal hygiene or social distancing. These actions in turn can lower their susceptibility to the disease. The interdependence of epidemics over its corresponding awareness results in a mutually interacting and competitive progression of these dynamical processes.  This interplay between the spread of epidemics due to persistent physical contacts and its corresponding awareness through virtual contacts has immensely contributed to the ever growing interest in this area of research \cite{Buno,Chen,Funck-1,Funck-2,Funck-3,kiss,sias,Wu,Zhang}. The understanding of such interacting processes can help in the comprehensive planning and implementation of immunization programmes\cite{Bo,Weng}.

   Multiplex networks are a natural extension of single layer networks that have emerged as conducive frameworks to study the dynamics of such interdependent and mutually competing processes\cite{multiplex-1,multiplex-2,multiplex-3,multiplex-4}. These networks represent a special category of multi-layered networks in which the different interactions among the same individuals are described over the respective layers. The different layers of multiplex networks which support these dynamical processes can have similar or distinct internal structure from one another. The study of coupled progression of epidemics along with its awareness over multiplex networks have provided valuable insights into the mechanism of interaction between these two dynamical processes. The results of these studies have further highlighted the effect of their mutual interactions upon the criticality of the disease outbreak and the final infected fraction of individuals. In this regard, Sahneh \textit{et al.} have shown that the resilience of the agents’ population to the spread of disease can be enhanced by dissemination of information on another network\cite{Sahneh}. They also found an optimal information dissemination metric for different topologies. Granell \textit{et al.} have abstracted epidemic processes coexisting with awareness spreading over multiplex networks. Their study investigates the competing effects of the spreading of both awareness and the epidemic\cite{Granell-1}. In a subsequent work, their model has been generalized by relaxing the earlier assumptions, wherein infection by the epidemic implied immediate awareness and this awareness implied total immunization of the epidemic. The effect of mass broadcast of awareness (mass media) on the epidemic dynamics has also been included in this work \cite{Granell-2}. Wang \textit{et al} have studied the asymmetrically interacting spreading dynamics based on a susceptible-infected-recovered (SIR) model in multiplex networks\cite{wang}. In their work, it has been shown that the outbreak of disease can lead to the propagation of information, and a rise in the epidemic threshold. Jia-Qian Kan \textit{et al.}  following a multiplex based approach have studied the effects of awareness diffusion and self-initiated awareness behavior on epidemic spreading\cite{Jia}. Scata \textit{et al.} have considered the impact of heterogeneity and awareness in their model to describe the spread of epidemic on multiplex networks\cite{Scata}. Wang \textit{et al}  have studied the use of information diffusion in suppressing the spread of disease on multiplex networks\cite{wei}.

    In the above mentioned studies, the transmission rates for the spread of epidemics (${\bf \beta^I}$) along with its corresponding awareness rate (${\bf \beta^A}$) have been assumed to be stationary in time. The criticality of the outbreak and the subsequent evolution of these processes have been associated with the largest eigenvalue of the network adjacency matrix\cite{eigen-1,eigen-2,eigen-3,sagar}. From the results of these earlier studies it is known that, there is no outbreak of the disease when the rate of transmission is less than a critical value (i.e. ${\bf \beta^I\;<\;\beta_{c}}$). The infection probability of each node in this case is reduced to zero. The disease spreads when the rate of transmission is greater than or equal to the critical value (i.e. ${\bf \beta^I \geq \beta_{c}}$). For the latter cases, the fraction of individuals in the infected state is non-zero. Beyond the critical value ${\bf \beta_{c}}$, the fraction of infected individuals increases gradually with a further increase in the value of transmission rate (${\bf \beta^I}$). However, in real life one can find many instances where the rates of transmission may not be stationary but can vary in time\cite{beta-1}. The time variation of transmission rates may arise due to the inherent recurring nature of the epidemics with leading and recovering phases \cite{beta-2,beta-3,beta-4,beta-5,beta-6}. The leading or growing phase of the disease in such cases is associated with the increase in the rate of transmission ${\bf \beta}$, whereas it decreases in the recovery phase. Alternatively, the temporal variation in the rates of transmission may also arise when the time rate of spread of disease information is comparable to rate of spread of disease. In this case, in the absence of initial disease related information there is no correlation between the disease and its associated awareness to start with. Consequently the disease spreads at the maximum rate. At later times, the spread of epidemic causes an increase in the rates of transmission of awareness which results in the development of a correlation between the two processes. The spread of awareness resulting from the collective behavior of the nodes beyond the threshold value of the  awareness transmission rate lowers the rate of transmission of the disease. Such a scenario is in accordance with everyday observations whereby the epidemic initially spreads in a network (e.g. social or computer networks) in the absence of initial awareness. However, the spread of infection results in the development/availability of preventive measures (vaccines or anti-virus software) which reduces the infection level. The time variation of the transmission rates due to the co-evolution of these processes can modify the rate of spread of disease and its corresponding awareness. This may also simultaneously affect the threshold levels of these processes and the fraction of infected individuals. In general, the time variation of the respective rates of transmission for such co-evolving processes can have very complicated functional forms with an implicit or explicit dependence upon multiple factors.  The intensity and mechanism of temporal modulation of transmission rates may depend upon the topology of network which is manifested in terms of intra-layer as well as inter-layer linking pattern of nodes. Another important factor to be considered while taking the time variation of the rates is an assessment of the inter-dependency of the respective rates of transmission upon each other. Thus, it is evident that the introduction of time variation in the rates of transmission can provide new perspectives in the study of co-evolution of such interdependent processes. Consequently, it is a challenging problem to design a comprehensive mathematical framework for examining the cumulative effects of time variation of respective rates of transmission on the infection and awareness probabilities of the participating nodes.

   The present work aims to develop a novel framework to study the co-evolution of epidemics and its corresponding awareness over a multiplex network taking into account the time variation of respective rates of transmission. In this framework, the differential rate of transmission for each of the nodes on the epidemic layer corresponding to the spread of disease depends upon the awareness probability of the node on the awareness layer. Similarly, the differential rate of transmission for the spread of disease related awareness corresponding to each of the nodes on the awareness layer depend upon its infection probability on the awareness layer. The collective dynamics of the nodes over the different layers of multiplex network results in the co-propagation of the epidemics and its corresponding awareness which in turn causes the temporal modulation of the transmission rates. The extent of temporal modulation of the differential rates of transmission depends upon both intra as well as inter layer linking patterns of the nodes. Thus, the new framework is self-consistently able to encompass the role of network topology in the temporal modulation of respective rates of transmission as well as upon the infection and awareness probabilities of the nodes. The spread of infection in this framework is described in terms of a non-linear Susceptible Infected Susceptible process (SIS) on the disease layer of a multiplex network. The corresponding spread of awareness induced by disease outbreak is represented as a cyclic process over another layer that undergoes a cycle of unaware-aware-unaware (UAU) states. A discrete time Monte Carlo Markov chain method is used to describe the spread of disease and its corresponding awareness. Using a second order linear theory, it is shown that in the continuous time limit, the progression of infection as well as awareness probabilities can be described in terms of coupled damped and driven oscillator equations. From the results of theoretical and Monte Carlo simulations over a scale free multiplex network, it is shown that the rates of transmission for each of these processes are not uniformly modulated in time. The modulation intensity of these transmission rates exhibit a joint dependence upon the degree of individual nodes in each of the multiplex layers as well as on the relative difference between them.  Moreover, the time variation of respective rates of transmission for each of these aforementioned dynamical processes co-evolving over different layers are found to have mutual dependencies upon each other. It is also found that, there is an inherent asymmetry in the spread of disease and its associated awareness over the respective layers. Further, the infection probability of nodes has a simultaneous dependence on the level of awareness as well as on the absolute and relative degrees of linkages.

 The organization of the paper is as follows:  In section (II), a theoretical frame work for studying the temporal co-evolution of disease and its associated awareness over the finite sized scale free network is presented.  In section (III), the mathematical formulation is described for the spread of co-evolving mutually interacting processes. In section (IV), the results of the Monte-Carlo simulations are presented highlighting the interdependence of various parameters together with the network topology upon outbreak, prevalence and size of epidemic.  Section (IV), contains a summary and discussion of the results.
          
\section{Theoretical Framework}
 A theoretical framework is described in this section to study the co-evolution of mutually interacting dynamical processes over a two layered multiplex network. The spread of infection due to the disease is described on the ${L_{1}}$ layer and its corresponding information induced awareness on the ${L_{2}}$ layer of the multiplex network. The multiplex network can be constructed by a one-to-one matching of the nodes in each layer with a specific set of rules. These rules for inter-linking the nodes determine the degree of correlation between the network layers\cite{wang}. For the uncorrelated networks, the degree distribution in one of the layers is completely independent of the distribution in the other layer. The hub node in this case with a high degree of links in one of the layers may not be matched to the hub in another layer. On the contrary, the matching of inter-layer nodes with similar intra-layer degree results in a correlated multiplex network. The focus of the present work is on multiplex networks that support interdependent dynamical processes such as disease and its corresponding awareness. For such multiplex networks, there is often some positive correlation in the mapping of inter-layer nodes\cite{wang}. This corresponds to the realistic case, where an ‘‘important’’ individual with high degree of links in one of the layers (e.g., representing persistent physical interactions) tends to have similar linking pattern on another network layer that reflect other kind of relations among the same individuals (e.g., representing virtual interactions). It is also worthwhile to mention here, that the earlier studies have highlighted the dependence of rate of spread of dynamical variable viz. disease, information etc. over complex network with fixed number of nodes upon the number of links\cite{now,sagar}. Thus in order to avoid the additional effects arising due to difference in layer density, the ratio of total number of nodes to total number of links has been kept the same and each of the layers differ only in their linking patterns.

  The multiplex networks describing the different modes of interactions within a population can be suitably described in terms of the corresponding graphs. In this description, the individuals are represented in terms of the nodes and the mutual connections (or links) between them as edges. The adjacency matrices $A_{L}=[a_{ij}]_{N\times N}$ and $B_{L}=[b_{ij}]_{N\times N}$  for each of these graphs are binary valued matrices in which the presence or absence of connection between a pair of nodes in each layer $a (i,j)$ and $b (i,j)$ is marked by numerical values of one (1) and zero (0) respectively. The edges (or connections) between the N nodes (or individuals) forming each of the multiplex layers are described in terms of a simple, un-weighted and undirected graph with a power law degree distribution $P (k)\sim k^{-\gamma}$ (where $\gamma >0$) based upon Barab\`{a}si-Alberta (BA) Model\cite{BA-1,BA-2}.

  In the present work, the concurrent spread of disease and its information induced awareness over the two layered multiplex network is based on a prototype {\bf SIS-UAU} model. It belongs to the class of compartmental models in which the individuals are partitioned in different compartments or states depending upon the stage of infection as well as their level of awareness. In this multiplex configuration, an individual in the disease layer (${L_{1}}$) can be in either of the two states namely susceptible (denoted by {\bf S}, those which are prone to infection) or infected (denoted by {\bf I}, those which are carrying the infection and are thus contagious). Similarly, an individual on the awareness layer (${L_{2}}$) can also be in either of two states, namely, unaware (denoted by {\bf U}, those which are ignorant of the presence of disease) or aware (denoted by {\bf A}, those which have the knowledge of the disease). The model follows a paradigmatic approach in which the microscopic details for the cause of infection as well as its resultant awareness are ignored and the states are assessed only macroscopically. The  individuals are neither assumed to have permanent immunity to the disease nor do they have persistent memory about it, which results in the loss of awareness. The individuals on the disease layer undergo stochastically looped over transitions from being initially susceptible (${\bf S}$) to getting infected ({\bf I}) by the disease at an awareness dependent rate $\beta^{I}(P^{A}(t))$ and then recovering from the infection to be again susceptible ({\bf S}) to it at rate ${\mu^{I}}$. Alternately, the individuals on the awareness layer undergo closed cycle of stochastic transitions from being initially unaware (or ignorant {\bf U}) to getting informed (or aware {\bf A}) resulting in the disease awareness at infection dependent rate ${\beta^{A}(P^{I}(t))}$ and then losing awareness to be again unaware (or ignorant {\bf U}) of it at a rate ${\mu^{A}}$. The state of an individual in this framework is jointly expressed in terms of its infection (${P^{I}_{i}}$) and awareness (${P^{A}_{i}}$) marginal probabilities. The schematic layout for the co-evolution of these processes over the respective layers of the multiplex network is described in Fig.~1.

    The mechanism for the spread of the disease along with its information on the infection and awareness layers of the multiplex can be understood in the following way: At an initial time instant which corresponds to the disease free state, the individuals on the respective layers of the multiplex network are in the susceptible ({\bf S}) and unaware ({\bf U}) states. In the absence of awareness (i.e {\bf U} ), the disease spreads on the infection layer (${L_{1}}$) with maximum rate of transmission ($\beta^I(0)=1$). The infection of the $i^{th}$ node on the physical layer ($L_{1}$) increments the value of its corresponding transmission rate ($\beta^A_{i}>0$) on the virtual layer($L_{2}$). In the successive steps more and more nodes are infected and thereby becomes fully aware. In other words there is a finite value of the awareness transmission rate. The spread of the awareness is a result of the collective behavior of the nodes on the virtual layer ($L_{2}$) whereby the leading eigenvalue of the transition matrix corresponding to these transmission rates becomes greater than or equal to one. Thus it is evident that the spread of disease information on the awareness layer is a result of cumulative effect of self-awareness due to infection on the disease layer (${L_{1}}$) and the interaction among neighboring nodes on the awareness layer (${L_{2}}$). The gradual increase in the awareness probability (${P^{A}_{i}}$) in turn inhibits the rate of progression of the disease by lowering the disease transmission rates (${\beta^{I}_{i}}$) on the infection layer (${L_{1}}$). The reduction in the disease transmission rate enhances the recovery process which thereupon lowers the infection probability of the individuals on the infection layer (${L_{1}}$). The decrease in the infection probability is often accompanied by the decline in the preventive measures, which in turn retards the rate of spread of corresponding awareness on the information layer (${L_{2}}$). On recovering from the infection, the individuals become susceptible to the disease and tend to lose awareness. This loss of awareness can again increase the rate of transmission of the disease on the infection layer and thus prompt a repeat of the whole process. Such a coupling between two mutually competing processes on the multiplex layers is analogous to the inductive coupling in the electric circuits in which the buildup or decline of current is self-consistently opposed by the resulting electromotive force (E.M.F) and vice versa.

\section{Mathematical Formulation}
The non-linear discrete time mathematical framework is described in this section to study the temporal evolution of aforementioned interdependent dynamical processes. The probability tree in Fig. 2, depicts at each time step, the possible states and their transition on the respective layers of the multiplex network. The spread of infection is described in terms of SIS ({\bf S}susceptible$\rightarrow${\bf I}nfected$\rightarrow${\bf S}uceptible) process in which the individual nodes can be either in the susceptible (${\bf S}$) or in the infected state (${\bf I}$). The discrete-time evolution of marginal probabilities of the nodes on the infection layer (${L_{1}}$) of the multiplex can be expressed as,
\begin{eqnarray}
 {P^{S}_{i}}(t + 1) &=& {Q^{I}_{i}}(t) {P^{S}_{i}}(t) + \mu^{I}{P^{I}_{i}}(t)Q^{I}_{i}(t), \label{eq-1}\\
 {P^{I}_{i}}(t + 1) &=& (1-{Q^{I}_{i}}(t)) {P^{S}_{i}}(t) + (1-\mu^{I}){P^{I}_{i}}(t)+{P^{I}_{i}}(t)\mu^{I}(1-{Q^{I}_{i}}(t)), \label{eq-2} \\
  {Q^{I}_{i}}(t) &=&\prod_{j=1}^{N}[1-{\beta^{I}_{i}}({P^{A}_{i}}(t)){a}_{ji} {P^{I}_{j}}(t)]. \label{eq-3}
\end{eqnarray}
In the above equation ${P^{S}_{i}}(t)$ and ${P^{I}_{i}}(t) $  are the probabilities for the $i^{th}$ node on the infection layer $L_{1}$ to be in the states of {\bf S} and {\bf I} at time t, respectively; ${Q^{I}_{i}}(t)$ represents the probability of $i^{th}$ node in the infection layer of not being infected by any of the neighbors. In the above expression, $\mu^{I}$ is the rate of recovery on the infection layer. The role of the individual terms in  Eq.(\ref{eq-2}) can be understood from the model description outlined in the following references\cite{gomez-1,gomez-2}. The first term corresponds to the probability that a susceptible node $(1-P^{I}_{i}(t))$ is infected $(1-Q^{I}_{i}(t))$ by at least one neighbor.  The second term gives the probability that the node infected at time $t$ does not recover $(1-\mu^{I})P^{I}_{i}(t)$, and finally the last term takes into account the probability that an infected node recovers $(\mu^{I} P^{I}_{i}(t))$ but is re-infected by at least a neighbor $(1-Q^{I}_{i}(t))$.  They satisfy the conservation condition :${P^{I}_{i}(t+1)+P^{S}_{i}(t+1)}={P^{I}_{i}(t)+P^{S}_{i}(t)}=1$. This implies that the probability of a node to be in the susceptible state can be expressed in terms of its infection probability i.e. ${P^{S}_{i}}(t) = 1-{P^{I}_{i}}(t) $. The conservation condition can be used to reduce the set of Eqs.({\ref{eq-1}-\ref{eq-2}}) to a single equation by expressing ${P^{S}_{i}}(t)$ in terms of ${P^{I}_{i}}(t)$.
\begin{eqnarray}
 {P^{I}_{i}}(t + 1) &=& (1-{Q^{I}_{i}}(t)) (1-{P^{I}_{i}}(t)) + (1-\mu^{I}){P^{I}_{i}}(t)+{P^{I}_{i}}(t)\mu^{I}(1-{Q^{I}_{i}}(t)). \label{eq-4}
\end{eqnarray}

 The spread of disease information induced awareness is described in terms of UAU ({\bf U}naware$\rightarrow${\bf A}ware$\rightarrow${\bf U}naware) process on information layer ($L_{2}$)in which the individual nodes can be either in an unaware (${\bf U}$) or an aware state (${\bf A}$). The discrete-time evolution of marginal probabilities of the nodes on awareness layer (${\bf L_{2}}$) of the multiplex can be expressed as,
\begin{eqnarray}
 {P^{U}_{i}}(t + 1) &=& {Q^{A}_{i}}(t) {P^{U}_{i}}(t) + \mu^{A}{P^{A}_{i}}(t)Q^{A}_{i}(t), \label{eq-5} \\
 {P^{A}_{i}}(t + 1) &=& (1-{Q^{A}_{i}}(t)) {P^{U}_{i}}(t) + (1-\mu^{A}){P^{A}_{i}}(t)+\mu^{A} (1-{Q^{A}_{i}}(t)){P^{A}_{i}}(t), \label{eq-6} \\
 {Q^{A}_{i}}(t) &=&\prod_{j=1}^{N}[1-{\beta^{A}_{i}}({P^{I}_{i}}(t)){b}_{ji} {P^{A}_{j}}(t)]. \label{eq-7}
\end{eqnarray}

In the above equation ${P^{U}_{i}}(t)$ and ${P^{A}_{i}}(t) $  are the probabilities for the $i^{th}$ node on the awareness layer $L_{2}$ to be in the states of {\bf U} and {\bf A} at time t, respectively;${Q^{A}_{i}}(t)$ represents the probability of $i^{th}$ node in the awareness layer of not being informed by any of the neighbors. In the above expressions $\mu^{A}$ is the rate at which a node loses its awareness. The role of the individual terms contained in  Eq.(\ref{eq-6}) that describes the dynamics of awareness on layer ${\L_{2}}$, can be understood in the following way. The first term defines the probability that an unaware node $(1-P^{U}_{i}(t))$ is informed $(1-Q^{A}_{i}(t)$ by at least a neighbor. The second term gives the probability that an informed node at time $t$ remains aware of the disease $(1-\mu^{A})P^{A}_{i}(t)$, and finally the last term takes into account the probability that an informed node loses its awareness $(\mu^{A} P^{A}_{i}(t))$ but is re-informed by at least a neighbor $(1-Q^{A}_{i}(t))$. They satisfy the conservation condition :${P^{A}_{i}(t+1)+P^{U}_{i}(t+1)}={P^{A}_{i}(t)+P^{U}_{i}(t)}=1$. This implies that the probability of node to be in the aware state can be expressed in terms of its unawareness probability i.e. ${P^{A}_{i}}(t) = 1-{P^{U}_{i}}(t) $. The conservation condition can be used to reduce the set of Eqs.({\ref{eq-5}-\ref{eq-6}}) to a single equation by expressing ${P^{U}_{i}}(t)$ in terms of ${P^{A}_{i}}(t)$.
\begin{eqnarray}
 {P^{A}_{i}}(t + 1) &=& (1-{Q^{A}_{i}}(t)) ({1-P^{A}_{i}}(t)) + (1-\mu^{A}){P^{A}_{i}}(t)+\mu^{A} (1-{Q^{A}_{i}}(t)){P^{A}_{i}}(t). \label{eq-8}
\end{eqnarray}

 In the above set of equations Eq.(\ref{eq-3} and \ref{eq-7}), ${\beta^{I}_{i}}({P^{A}_{i}}(t))$ and ${\beta^{A}_{i}}({P^{I}_{i}}(t))$ are awareness and infection probability dependent rates of transmission of disease and its corresponding awareness. The infection (${\beta^{I}_{i}}({P^{A}_{i}}(t))$) and awareness (${\beta^{A}_{i}}({P^{I}_{i}}(t))$) transmission rates are normalized to the respective rates of recovery ($\mu^{I}$) and rate of loss of awareness ($\mu^{A}$).The dependence described here is general and the explicit mathematical form can be chosen in accordance with the requirement of the problem. In the present work, a linear dependence has been assumed to describe the inductively coupled rates of the transmission on each layer. These rates can be given by,
\begin{eqnarray}
  {\beta^{I}_{i}}({P^{A}_{i}}(t)) &=& \beta^{I}_{i_{0}}(1-{\alpha^{I}_{i}} {P^{A}_{i}}(t)),  \label{eq-9} \\
  {\beta^{A}_{i}}({P^{I}_{i}}(t)) &=& \beta^{A}_{i_{0}}(1-{\alpha^{A}_{i}} (1-{P^{I}_{i}}(t))). \label{eq-10}
\end{eqnarray}
 From Eq.\ref{eq-9} and Eq.\ref{eq-10} it is evident that the maximum and minimum values of the temporally modulated transmission rates for the spread of infection as well as information depend upon the state of nodes in the opposite layers of the multiplex network. For the $i^{th}$ node in the multiplex network, the maximum probable state in both the process corresponds to a numerical value of unity i.e. $P_{i}^I =1$ and $P_{i}^{A}=1$ and the minimum value is zero i.e. $P_{i}^I =0$ and $P_{i}^{A}=0$. The maximum in the value of awareness and infection probabilities results in minimum value for the infection transmission rate (i.e. $ \beta_{i_{min}}^{I} =\beta^{I}_{i_{0}}(1-{\alpha^{I}_{i}})$) and maximum for the corresponding information transmission rates (i.e. $\beta^{A}_{i_{max}}=\beta^{A}_{i_{0}}$) respectively. On the contrary, the minimum value of awareness and infection probabilities results in the maximum infection transmission rate (i.e. $\beta^{I}_{i_{max}}=\beta^{I}_{i_{0}}$) and the minimum awareness transmission rate (i.e. $ \beta_{i_{min}}^{A} =\beta^{A}_{i_{0}}(1-{\alpha^{A}_{i}})$). The coefficients $\alpha^{I}_{i}$ and $\alpha^{A}_{i}$ in the equations Eq.(\ref{eq-9}-\ref{eq-10}) couple the participation of the individual nodes in the dynamical processes on the different layers of the multiplex network. In this framework, the choice of $\alpha^{I}_{i} = \alpha^{A}_{i}=0$  decouples the spread of mutually coupled dynamical processes on each of the multiplex layer. The role of these coefficients is further elaborated in the subsequent sub-section. The parameters  $\beta^{I}_{i_{0}}$ and $\beta^{A}_{i_{0}}$, define the amplitudes of rates of transmission of infections and awareness of $i^{th}$ nodes on the respective layers of multiplex network.

The time averaged probability density of each layer is given by,
\begin{eqnarray}
  {\rho^{I}_{A}}_{av}(t) &=&\frac{1}{N} \sum_{j=1}^{N}P^{I}_{j}(t), \label{eq-11}\\
  {\rho^{A}_{B}}_{av}(t) &=&\frac{1}{N} \sum_{j=1}^{N}P^{A}_{j}(t). \label{eq-12}
\end{eqnarray}

\subsection{Discrete Time Linear Model: Choice of Initial Conditions}

The interactions between the disease and information spreading processes described by Eq.\ref{eq-1}-\ref{eq-10} have a complicated mathematical form. Thus it is not feasible to analytically estimate the exact threshold values. Therefore, in accordance with the procedure outlined in references 25, 28 and 36, the critical properties of the system can be estimated by linearizing the $N$-dimensional nonlinear stochastic system described by equations Eqs.(\ref{eq-4} and \ref{eq-8}). Using Eq.(\ref{eq-3} and \ref{eq-7}) about the unaware disease free state given by $P^{I}_{i}=0;P^{A}_{i}=0$, to the first order and removing all higher order terms, leads to,
\begin{eqnarray}
  {Q^{I}_{i}}(t) &\approx&1-{\beta^{I}_{i}}({P^{A}_{i}}(t))\sum_{j=1}^{N}{a}_{ji} P^{I}_{j}(t). \label{eq-13} \\
  {Q^{A}_{i}}(t) &\approx&1-{\beta^{A}_{i}}({P^{I}_{i}}(t))\sum_{j=1}^{N}{b}_{ji} P^{A}_{j}(t). \label{eq-14}
\end{eqnarray}
The resultant upper bound linear approximate model corresponding to the nonlinear model described by Eq.\ref{eq-2} and Eq.\ref{eq-4} can be expressed as,
\begin{eqnarray}
 {P^{I}_{i}}(t+1) &=& (1-\mu^{I}){P^{I}_{i}}(t)+{\beta^{I}_{i}}({P^{A}_{i}}(t))\sum_{j=1}^{N}{a}_{ji} {P^{I}}_{j}(t)=\sum_{j=1}^{N}{m_{I}}_{ij}(t)P^{I}_{j}(t), \label{eq-15} \\
 {P^{A}_{i}}(t+1) &=& (1-\mu^{A}){P^{A}_{i}}(t)+{\beta^{A}_{i}}({P^{I}_{i}}(t))\sum_{j=1}^{N}{b}_{ji} {P^{A}}_{j}(t)=\sum_{j=1}^{N}{m_{A}}_{ij}(t)P^{A}_{j}(t). \label{eq-16}
\end{eqnarray}
In the above equations no difference has been assumed in the rate of recovery and loss of information of an individual (i.e.$\mu^{I}$ and $\mu^{A}$ are constant). The equation Eq.\ref{eq-15} and Eq.\ref{eq-16} can be expressed in matrix form by representing $P^{I}_{i}(t)$ and $P^{A}_{i}(t)$ as $N$-dimensional column vector obtained from the marginal probability $P^{I}_{i}(t)$ and $P^{A}_{i}(t)$,
\begin{eqnarray}
P^{I}(t+1)&=& M_{1}(t)P^{I}(t), \label{eq-17}\\
P^{A}(t+1)&=& M_{2}(t)P^{A}(t). \label{eq-18}
\end{eqnarray}

where, $M_{1}(t)=\left[ {m_{1}}_{ij} \right]$, ${m_{1}}_{ij}={\beta^{I}_{i}}({P^{I}_{i}}(t)){a}_{ij}+{\Delta_I}_{ij}(1-\mu^{I})$ and ${\Delta_{I}}_{ij}$ is the Kronecker delta corresponding to the infection layer ($L_{1}$). Similarly, $M_{2}(t)=\left[ {m_{2}}_{ij} \right]$, ${m_{2}}_{ij}={\beta^{A}_{i}}({P^{A}_{i}}(t)){b}_{ij}+{\Delta_A}_{ij}(1-\mu^{A})$ and ${\Delta_{A}}_{ij}$ is the Kronecker delta corresponding to the awareness layer ($L_{2}$). These expressions can be re-written using transmission rates for Eq.\ref{eq-9} and Eq.\ref{eq-10} : $M_{1}(t)=\left[ {m_{1}}_{ij} \right]$, ${m_{1}}_{ij}=\beta^{I}_{i_{0}}(1-{\alpha^{I}_{i}} {P^{A}_{i}}(t)){a}_{ij}+{\Delta_I}_{ij}(1-\mu^{I})$ and $M_{2}(t)=\left[ {m_{2}}_{ij} \right]$, ${m_{2}}_{ij}=\beta^{A}_{i_{0}}(1-{\alpha^{A}_{i}} (1-{P^{I}_{i}}(t))){b}_{ij}+{\Delta_A}_{ij}(1-\mu^{A})$. It is known from previous studies\cite{eigen-1,eigen-4,sagar}, that for both homogeneous and heterogeneous rates of transmission the disease free state is asymptotically stable if the largest eigenvalue of $M$ given by $\zeta_{M}$ satisfies the condition $\zeta_{M}<1$. Thus for $\zeta_{M}<1$, the infection or awareness will die out exponentially fast with the rate determined by $\zeta_{M}$. The present study is intended to investigate the effect of time variation of the transmission rates resulting from their mutual coupling. For this a numerical investigation has been carried out for the following three cases: In the first case, the value of structure functions is chosen such that the minimum critical rates (i.e.${\beta^{I}_{i_{min}}}$ and ${\beta^{A}_{i_{min}}}$) in each layer results in the eigenvalues of matrix $M_{1}$ and $M_{2}$ being less than one i.e.$\zeta_{M_{1}}<1$ and $\zeta_{M_{2}}<1$ for which their is no outbreak of dynamical process (i.e. infection or awareness). In the second case, the value of structure functions is chosen such that the minimum critical rates (i.e.${\beta^{I}_{i_{min}}}$ and ${\beta^{A}_{i_{min}}}$) in each layer results in the eigenvalues of matrix $M_{1}$ and $M_{2}$ being nearly equal to one i.e.$\zeta_{M_{I}} \approx 1$ and $\zeta_{M_{1}}\approx 1$ for which there is an outbreak of the dynamical process. In the third case, the value of structure functions is chosen such that the minimum critical rates (i.e.${\beta^{I}_{i_{min}}}$ and ${\beta^{A}_{i_{min}}}$) in each layer results in the eigenvalues of matrix $M_{1}$ and $M_{2}$ being greater than one i.e. $\zeta_{M_{1}} > 1$ and $\zeta_{M_{2}} > 1$.

\subsection{Continuous Time}
In the study of dynamical processes over complex networks, one is often interested in the average properties rather than individual dynamics. The averaging procedure requires running sufficiently many realizations of the discrete time stochastic model to produce such average results. The computational cost to obtain such multiple realizations increases with the increase in the size of networks together with the number of modes interaction between them. Therefore, deterministic versions of the stochastic models can be used to obtain the average dynamics of the spreading process and make predictions about its future state. The results from the deterministic model are expected to be comparable with those obtained from the averaged values of the stochastic realizations in the limit of large network sizes\cite{pone}. In addition to this, the use of deterministic version also allows for a simpler analysis of the dynamical behavior of the model a subject for consideration in future works. Thus at this time it is appropriate to consider the  dynamics of each layer in the continuous time limit and derive the corresponding evolution equations. On simplifying and re-writing Eq.(\ref{eq-4} and \ref{eq-8}) one can obtain,
\begin{eqnarray}
  {P^{I}_{i}} (t + \Delta t) &=& (1-{Q^{I}_{i}}(t)) + (1-\mu^{I}){P^{I}_{i}}(t){Q^{I}_{i}}(t),\label{eq-19} \\
  {P^{A}_{i}} (t + \Delta t) &=& (1-{Q^{A}_{i}}(t)) + (1-\mu^{A}){P^{A}_{i}}(t){Q^{A}_{i}}(t),\label{eq-20}
\end{eqnarray}
In the continuous limit ${\beta^{I}_{i}}({P^{A}_{i}}(t))=\hat{\beta}^{I}_{i}({P^{A}_{i}}(t))\Delta t$, ${\beta^{A}_{i}}({P^{I}_{i}}(t))=\hat{\beta}^{A}_{i}({P^{I}_{i}}(t))\Delta t$, the respective recovery rates take the form $\mu^{I}=\hat{\mu}^{I}\Delta t$ and $\mu^{A}=\hat{\mu}^{A}\Delta t$.
 On linearizing the probability,
\begin{eqnarray}
  {Q^{I}_{i}}(t) &\approx&\hat{\beta}^{I}_{i_{0}}(1-\hat{\beta}^{I}_{i}({P^{A}_{i}}(t))\Delta t \sum_{j=1}^{N}{a}_{ji}P^{I}_{j}(t)), \label{eq-21} \\
  {Q^{A}_{i}}(t) &\approx&\hat{\beta}^{A}_{i_{0}}(1-\hat{\beta}^{A}_{i}({P^{A}_{i}}(t))\Delta t \sum_{j=1}^{N}{b}_{ji}P^{A}_{j}(t)). \label{eq-22}
\end{eqnarray}
 For the continuous time limit, the time evolution of the marginal probability can be written as,
\begin{eqnarray}
\dot{P^{I}_{i}}=\text{lt}_{\Delta t\rightarrow 0}\frac{{P^{I}_{i}}(t+\Delta t)-{P^{I}_{i}}(t)}{\Delta t} &=& \frac{1}{\Delta t} \bigg(-\hat{\mu}^{I}\Delta t {P^{I}_{i}}(t)+\hat{\beta}^{I}_{i}({P^{A}_{i}}(t))\Delta t(1-{P^{I}_{i}}(t)) \sum_{j=1}^{N}{a}_{ji} P^{I}_{j}(t)+ \nonumber \\
&& \hat{\mu}^{I}\hat{\beta}^{I}_{i}({P^{A}_{i}}(t)){\Delta t}^2 {P^{I}_{i}}(t)\sum_{j=1}^{N}{a}_{ji} P^{I}_{j}(t)\bigg) \label{eq-23} \\
\dot{P^{A}_{i}}=\text{lt}_{\Delta t\rightarrow 0}\frac{{P^{A}_{i}}(t+\Delta t)-{P^{A}_{i}}(t)}{\Delta t} &=& \frac{1}{\Delta t}\bigg(-\hat{\mu}^{A}\Delta t {P^{A}_{i}}(t)+\hat{\beta}^{A}_{i}({P^{I}_{i}}(t))\Delta t(1-{P^{A}_{i}}(t)) \sum_{j=1}^{N}{b}_{ji} {P^{A}_{j}}(t)+\nonumber \\
&& \hat{\mu}^{A}\hat{\beta}^{A}_{i}({P^{I}_{i}}(t)){\Delta t}^2 {P^{A}_{i}}(t) \sum_{j=1}^{N}{b}_{ji} P^{A}_{j}(t)\bigg)  \label{eq-24}
\end{eqnarray}
By rearranging, the equations can be re-written as,
\begin{eqnarray}
 \dot{P^{I}_{i}}(t) &=& -\hat{\mu}^{I}{P^{I}_{i}}(t)+\hat{\beta}^{I}_{i}({P^{A}_{i}}(t))(1-{P^{I}_{i}}(t)) \sum_{j=1}^{N}{a}_{ji} {P^{I}}_{j}(t), \label{eq-25} \\
 \dot{P^{A}_{i}}(t) &=& -\hat{\mu}^{A}{P^{A}_{i}}(t)+\hat{\beta}^{A}_{i}({P^{I}_{i}}(t))(1-{P^{A}_{i}}(t)) \sum_{j=1}^{N}{b}_{ji} {P^{A}}_{j}(t).  \label{eq-26}
\end{eqnarray}
For the linear dependence the rate of transmission can be expressed as,
\begin{eqnarray}
  \hat{\beta}^{I}_{i}({P^{A}_{i}}(t)) &=& \hat{\beta}^{I}_{i_{0}}(1-\alpha^{I}_{i} {P^{A}_{i}}(t)), \label{eq-27} \\
  \hat{\beta}^{A}_{i}({P^{I}_{i}}(t)) &=& \hat{\beta}^{A}_{i_{0}}(1-\alpha^{A}_{i} (1-{P^{I}_{i}}(t))). \label{eq-28}
\end{eqnarray}
The substitution of the rate expression results in,
\begin{eqnarray}
 \dot{P^{I}_{i}}(t)&=&-\hat{\mu}^{I}{P^{I}_{i}}(t)+{\hat{\beta}^{I}_{i_{0}}}(1-\alpha^{I}_{i} {P^{A}_{i}}(t))(1-{P^{I}_{i}}(t))\sum_{j=1}^{N}{a}_{ji}P^{I}_{j}(t) \label{eq-29} \\
 \dot{P^{A}_{i}}(t)&=&-\hat{\mu}^{A}{P^{A}_{i}}(t)+{\hat{\beta}^{A}_{i_{0}}}(1-\alpha^{A}_{i} (1-{P^{I}_{i}}(t)))(1-{P^{A}_{i}}(t))\sum_{j=1}^{N}{b}_{ji}P^{A}_{j}(t)  \label{eq-30}
\end{eqnarray}
The above derived first order equations are coupled to the zeroth order time dependence of the probabilities. These equations can be decoupled by taking the second time derivative.
\begin{eqnarray}
\ddot{P^{I}_{i}}(t)&=&-\hat{\mu}^{I}\dot{P^{I}_{i}}(t)-{\hat{\beta}^{I}_{i_{0}}}\alpha^{I}_{i}\dot{P^{A}_{i}}(t)(1-{P^{I}_{i}}(t))\sum_{j=1}^{N}{a}_{ji} {P^{I}}_{j}(t)+{\hat{\beta}^{I}_{i_{0}}}(1-\alpha^{I}_{i} {P^{A}_{i}}(t))\nonumber \\
&& \bigg( (1-{P^{I}_{i}}(t))\sum_{j=1}^{N}{a}_{ji} {P^{I}}_{j}(t)\bigg)^{\prime}  \label{eq-31} \\
\ddot{P^{A}_{i}}(t)&=&-\hat{\mu}^{A}\dot{P^{A}_{i}}(t)+{\hat{\beta}^{A}_{i_{0}}}\alpha^{A}_{i}\dot{P^{I}_{i}}(t)(1-{P^{A}_{i}}(t)) \sum_{j=1}^{N}{b}_{ji} {P^{A}}_{j}(t)+{\hat{\beta}^{A}_{i_{0}}}(1-\alpha^{A}_{i}(1-{P^{I}_{i}}(t)))\nonumber \\
&&\bigg((1-P^{A}_{i}(t)) \sum_{j=1}^{N}{b}_{ji} P^{A}_{j}(t)\bigg)^{\prime}     \label{eq-32}
\end{eqnarray}

On substituting the value of $\dot{P^{I}_{i}}(t)$ and $\dot{P^{A}_{i}}(t)$, one gets,
\begin{eqnarray}
 \ddot{P^{I}_{i}}(t) &=& -\hat{\mu}^{I}\dot{P^{I}_{i}}(t)-{\hat{\beta}^{I}_{i_{0}}}\alpha^{I}_{i}\bigg(-\hat{\mu}^{A}{P^{A}_{i}}(t)+{\hat{\beta}^{A}_{i_{0}}}(1-\alpha^{A}_{i} (1-{P^{I}_{i}}(t)))(1-{P^{A}_{i}}(t))\sum_{j=1}^{N}{b}_{ji}P^{A}_{j}(t) \bigg)\nonumber \\
&& (1-{P^{I}_{i}}(t))\sum_{j=1}^{N}{a}_{ji} {P^{I}}_{j}(t)+{\hat{\beta}^{I}_{i_{0}}}(1-\alpha^{I}_{i} {P^{A}_{i}}(t))\bigg( (1-{P^{I}_{i}}(t))\sum_{j=1}^{N}{a}_{ji} {P^{I}}_{j}(t)\bigg)^{\prime}  \label{eq-33} \\ \nonumber
 \ddot{P^{A}_{i}}(t) &=& -\hat{\mu}^{A}\dot{P^{A}_{i}}(t)+{\hat{\beta}^{A}_{i_{0}}} \alpha^{A}_{i}\bigg(-\hat{\mu}^{I}{P^{I}_{i}}(t)+{\hat{\beta}^{I}_{i_{0}}}(1-\alpha^{I}_{i} {P^{A}_{i}}(t))(1-{P^{I}_{i}}(t))\sum_{j=1}^{N}{a}_{ji}P^{I}_{j}(t)  \bigg)\nonumber \\
 &&(1-{P^{A}_{i}}(t))\sum_{j=1}^{N}{b}_{ji} P^{A}_{j}(t)+{\hat{\beta}^{A}_{i_{0}}}(1-\alpha^{A}_{i}(1-{P^{I}_{i}}(t)))\bigg((1-{P^{A}_{i}}(t)) \sum_{j=1}^{N}{b}_{ji} P^{A}_{j}(t)\bigg)^{\prime}\label{eq-34}
\end{eqnarray}

The Eqs.\ref{eq-32} and \ref{eq-33} can be re-arranged and expressed in the form of damped driven oscillator equations,

\begin{eqnarray}
 \ddot{P^{I}_{i}}(t)+{\omega(t)}^{2}{P^{I}_{i}}(t)+\hat{\mu}^{I}\dot{{P^{I}_{i}}}(t)&=& F^{I}_{i}(t) \label{eq-35} \\
 \ddot{P^{A}_{i}}(t)+{\omega(t)}^{2}{P^{A}_{i}}(t)+\hat{\mu}^{A}\dot{{P^{A}_{i}}}(t) &=& F^{A}_{i}(t) \label{eq-36}
\end{eqnarray}

The coupled system oscillates at a frequency given by,
\begin{eqnarray}
{\omega(t)}^2&=& [\alpha^{I}_{i} \alpha^{A}_{i} \bigg( (1-{P^{A}_{i}}(t)) \sum_{j=1}^{N}{b}_{ji} P^{A}_{j}(t) \bigg) \bigg((1-{P^{I}_{i}}(t))\sum_{j=1}^{N}{a}_{ji} P^{I}_{j}(t)\bigg)
\end{eqnarray}
In the above equations, the rate of recovery ($\hat{\mu}^{A}$) and rate of loss of awareness ($\hat{\mu}_{B}$) correspond to damping coefficients. The self-consistent driving force arising from network structure dependent connection patterns can be expressed as,
\begin{eqnarray}
F^{I}_{i}(t)&=&-{\alpha^{I}_{i}\bigg( -\hat{\mu}^{A}}{P^{A}_{i}}(t)+ (1-\alpha^{A}_{i})(1-{P^{A}_{i}}(t)) \sum_{j=1}^{N}{b}_{ji} {P^{A}}_{j}(t)\bigg)(1-{P^{I}_{i}}(t))\sum_{j=1}^{N}{a}_{ji} P^{I}_{j}(t)+\nonumber \\
&&  (1-\alpha^{I}_{i} {P^{A}_{i}}(t))\bigg( (1-P^{I}_{i}(t))\sum_{j=1}^{N}{a}_{ji} P^{I}_{j}(t)\bigg)^{\prime} \\
F^{A}_{i}(t)&=& \alpha^{A}_{i}\bigg( -\hat{\mu}^{I} {P^{I}_{i}}(t)+(1-{P^{I}_{i}}(t))\sum_{j=1}^{N}{a}_{ji} P^{I}_{j}(t)\bigg)(1-{P^{A}_{i}}(t)) \sum_{j=1}^{N}{b}_{ji} P^{A}_{j}(t)+\nonumber \\
&&(1-\alpha^{A}_{i}(1-{P^{I}_{i}}(t)))\bigg((1-{P^{A}_{i}}(t)) \sum_{j=1}^{N}{b}_{ji} P^{A}_{j}(t)\bigg)^{\prime}
\end{eqnarray}
The method described above is quite general and can be applied to study a wide array of problems that are not just limited to social networks. As an example, the method can be applied to the study of computer networks that are susceptible to virus attacks. A detailed understanding of the mechanism of the spread of viruses can be used to device efficient immunization strategies.

\section{RESULTS}
 Our Monte Carlo simulation results are presented in this sub-section. They provide a detailed insight into the co-evolving dynamics of the disease along with its corresponding information over the infection (${L_{1}}$) and awareness (${L_{2}}$) layers by taking into account the self-consistent temporal modulations of the respective transmission rates. It is well known, that in the absence of such temporal variations, the outbreak and progression of these dynamical processes depend on the critical value of the respective transmission rates\cite{eigen-1,eigen-4,sagar}. The focus of the present work is twofold: firstly, to understand the mechanism of temporal modulation of these rates and to subsequently examine their effect on the critical properties of the aforementioned processes. The spread of disease in this framework is assumed to be caused by the persistent physical contacts and its dynamics is described over the infection layer (${L_{1}}$) of the multiplex network. The disease dynamics is based on the discrete time non-linear SIS model in which a given node at any time can be in either of the two states namely, susceptible ({\bf S}) or infected ({\bf I}). Similarly, the virtual contacts mediated spread of disease information is described over the awareness layer (${L_{2}}$) in terms of {\bf UAU} model for which each of the nodes can be in two states namely, unaware ({\bf U}) or aware ({\bf A}). The Barab\`{a}si-Albert (BA)\cite{BA-1,BA-2} model has been used to generate each layer of the scale free multiplex network with a power law degree distribution of the links. Each multiplex layer has the same number of nodes (N=$5\times10^2$) with a minimum of two links per node.

   Our study has been carried out in the parameter space of the variables ${{\alpha_{i}}_{[L_{1},L_{2}]}}$, which regulate the lower bound of the temporally modulated rates of transmission on the infection (${L_{1}}$) and awareness (${L_{2}}$) layers. These lower bounds correspond to their respective time invariant critical values for the outbreak and successive progression of each of these processes. The simulations describing the co-propagation of these processes have been divided in the following three parametric domains: In the first case, these variables (i.e.${{\alpha_{i}}_{[L_{1},L_{2}]}}$) are chosen such that lower bound of the modulated rates of transmission for each of these processes can be less than the respective time invariant critical values i.e. ${0<{{\beta^{I}_{min}}}< {\beta^{I}_{c}}}$ and ${0<{\beta^{A}_{min}} < {\beta^{A}_{c}}}$. In the second case, these variables (i.e.${{\alpha_{i}}_{[L_{1},L_{2}]}}$) are chosen such that the lower bound of the modulated rates of transmission for each of these processes is nearly equal to their respective critical values i.e. ${{{\beta^{I}_{min}}} \approx {\beta^{I}_{c}}}$ and  ${{{\beta^{A}_{min}}} \approx{\beta^{A}_{c}}}$. In the final case, the variables (i.e. ${{\alpha_{i}}_{[L_{1},L_{2}]}}$) are chosen such that the lower bound of the modulated rates of the transmission is greater than their critical value i.e. ${{{\beta^{I}_{min}}} > {\beta^{I}_{c}}}$ and ${{{\beta^{A}_{min}}} > {\beta^{A}_{c}}}$. At the onset of the simulation, a certain fraction of the nodes ($5\%$) are randomly infected. In the present work the transmission rates for the spread of infection and awareness are normalized to the respective rate of recovery ($\mu^{I}$) and loss of awareness ($\mu^{A}$). The rates of recovery for each of the processes are assumed to be unity (i.e.$\mu^{I}=1$ and $\mu^{A}=1$). The coupled set of equations given by Eqs.\ref {eq-4} and \ref {eq-8} describes the dynamical state of a network at any given time.

 The results of the parametric study are segregated in terms of three different cases (i.e. 1,2 and 3) which correspond to the aforementioned respective parametric domains. In the present work, the nodes supporting the co-progression of these dynamical processes in the multiplex layers for all the three cases are initially in the disease free (or susceptible {\bf S}) and unaware state ({\bf U}). The corresponding initial marginal probabilities for each of the layers and the respective transmission rates can be mathematically expressed as, ${{P^{I}_{i}}\approx 0}$ and ${{\beta^{I}_{i}}\approx 1}$  and  ${{P^{A}_{i}}\approx 0}$ and ${{\beta^{A}_{i}} \approx {{\beta^{A}_{i_{min}}}}}$. In all three cases, the absence of initial awareness results in the spread of disease on the infection layer (${L_{1}}$) at a maximum rate of transmission (${\beta^{I}_{i}}(0)\approx 1$).

 Fig.~3 describes the time evolution of infection ($P^{I}_{i}$) and awareness probabilities ($P^{A}_{i}$) for each of the multiplex nodes in the absence of any temporal modulation of the respective transmission rates. The temporally un modulated transmission rates for each of these processes on the infection ($L_{1}$) and awareness ($L_{2}$) layers of multiplex network can be expressed as: ${{\beta^{I}_{i}(t)=\beta^{I}_{i}(0)}=\beta_{i0}^{I}(1-\alpha_{i}^{I}P_{i}^{A}(0))}$ and ${{\beta^{A}_{i}(t)=\beta^{A}_{i}(0)}=\beta_{i0}^{A}(1-\alpha_{i}^{A}(1-P_{i}^{I}(0)))}$.  It is evident from the results described in case-{\bf 1}, that there is no outbreak and subsequent spread of awareness (i.e. $P_{i}^{A}=0$) on the layer $L_{2}$, when the awareness transmission rates modified by the initial infection probability are less than the corresponding critical value ( i.e $\beta_{i_{min}}^{[I,A]} < \beta_{i_{c}}^{[I,A]}$ ). Furthermore, the initial awareness (i.e. $P_{i}^{A}(0)$) modulated infection transmission rates are found to have no significant effect on the infection dynamics over the disease layer ($L_{1}$) and the infection probability saturates to maximum value (i.e. $P_{i}^{I}=1$). For the case-{\bf 2}, the values of parameters $\alpha_{i}^{A}$ have  been chosen such that the minimum transmission rates on the awareness layer ($L_{2}$) are comparable to the corresponding critical values ( i.e $\beta_{i_{min}}^{[I,A]}\approx \beta_{i_{c}}^{[I,A]}$ ). It is evident from the results that in this case there is an outbreak of the process resulting in a finite value of the awareness marginal probabilities (i.e. $P_{i}^{A}\neq 0)$. For the case-{\bf 3}, the increase in the transmission rates beyond critical values ( i.e $\beta_{i_{min}}^{[I,A]}> \beta_{i_{c}}^{[I,A]}$ ) increases the respective awareness probability. From the figure, it can be further inferred that, for both cases, {\bf 2} and {\bf 3}, the increase in awareness level ($P_{i}^{A}\neq 0$) in the absence of any temporal modulations, neither alters the rate of spread of infection nor reduces the infection probability which saturates to maximum value ($P^{I}_{i}=1$).

   Fig.~4 elucidates the effect of self-consistent temporal modulation of the respective rates of transmission for each of the three aforementioned cases. In comparison to the unmodulated case, in the presence of temporal modulation the infected individuals (i.e. ${P^{I}_{i} \neq 0}$) gain information about the disease which results in the increase of the transmission rates (${\beta^{A}_{i}}$) for these nodes on the awareness layer (${L_{2}}$). The increase in the information transmission rate in turn has a dual impact on the collective dynamics of the nodes in each of the layers: firstly, it triggers the spread of disease information on the awareness layer (${L_{2}}$); secondly, the spread of the information induced awareness tends to simultaneously retard the spread of disease on the other layer (${L_{1}}$) by lowering the corresponding transmission rates. The resultant reduction in the disease transmission rate aids in the process of recovery and thus lowers the infection probability (${{P^{I}_{i}}}$). The recovered nodes (or individuals) tend to lose the disease awareness and become susceptible to it. From the figure, it is evident that there is a marked difference between the progression characteristics of these processes corresponding to the case-{\bf (1)} and that of the cases-{\bf (2)} and {\bf (3)}.

   For case-{\bf 1}, this results in a repetitive cycle similar to the initial one, but due to an asymmetric spread of disease, the level of awareness is relatively enhanced in comparison to its initial value. The finite awareness lowers the disease transmission rate which reduces the peak infection probability of the nodes in the subsequent cycles. It can also be seen, that unlike the initial time instant, the rates of disease transmission for different nodes of the network are neither equal to unity nor do they have uniform values. The non-uniform modulation intensity of initially uniform transmission rates result in the variable levels of infection and its corresponding awareness. Furthermore, for this case, the system exhibits no endemic (or equilibrium) state and there are self-sustained alternating cycles of infection followed by awareness. Moreover, in comparison to the time independent case, there is an outbreak as well as a spread of disease even when the lower bound transmission rates are less than the critical value.

  For the cases {\bf (2) and (3)}, which correspond to the condition: ${\beta^{[I,A]}_{i_{min}} \geq \beta^{[I,A]}_{i_{c}}}$ , it can be seen, that the initial cycle describing the spread of disease and its resultant awareness over the respective layers of the multiplex is similar to that of case-{\bf (1)}. However, in subsequent cycles, the marginal probabilities for the infection and its corresponding awareness along with their respective transmission rates exhibit damped oscillatory progression instead of the sustained oscillatory progression. This indicates, that for these processes, the minimum rate of transmission greater than or equal to its critical value results in simultaneous prevalence of the infection and its corresponding awareness upon the respective layers of the multiplex. It is also evident that, the rate as well as the amplitude of the oscillation damping is not uniform for the respective transmission rates and marginal probabilities on either of the multiplex layers. On comparing the results for case {\bf (2)} and {\bf (3)}, it can be further observed that, the rate of amplitude damping tends to increases with increase in the lower bound of the respective rates of transmission. Moreover, the dynamical processes corresponding to these cases are associated with endemic (or equilibrium) state in which there is a finite probability for the disease to coexist with its resultant awareness (i.e. $ {{P^{I}_{i}} \neq 0, {P^{I}_{i}}(t+1)\approx{P^{I}_{i}}(t)}$ and ${{P^{A}_{i}}\neq 0,{P^{A}_{i}}(t+1)\approx{P^{A}_{i}}(t)}$). Thus from these results it can be concluded, that the presence of disease awareness can substantially lower both the rate of disease transmission and the marginal infection probability of the nodes. In addition, it is also found that the lower rate of disease transmission resulting from high awareness may not necessarily result in a low infection probability.

  The temporal evolution of the probabilities of the average infection $\rho_{avg}^{I}$ and awareness $\rho_{avg}^{A}$ as functions of their respective rates of transmissions  $\beta_{avg}^{I}$ and $\beta_{avg}^{A}$ are shown in Fig 5. This figure describes the phase relationship between the respective infection and awareness probabilities with their respective rates of transmissions as functions of the parameter $\alpha$ which in turn determines the lower bound of the transmission rates. It is evident from the figure that for case-({\bf 1}) which corresponds to the condition when the minimum individual rates of transmission are less than the threshold value the system exhibits sustained oscillations. Cases ({\bf 2}) and ({\bf 3}) correspond to the parametric regime in which the minimum rates of transmission are equal to or greater than the threshold value. In these regimes the system exhibits damped oscillations and for each of these cases the average probabilities along with their respective transmission rates attain a fixed value.

Detailed results of the simulations as shown in Fig.~4, provide insights into the mechanism of spread of disease and its corresponding awareness over the different layers of the multiplex network. The role of network topology on the co-evolution of these processes can be further understood by examining the dependence of the respective probabilities and their associated transmission rates on the intra-layer and inter-layer linking patterns.

  Fig.~6, describes the intra-layer degree based temporal evolution of infection and awareness probabilities (i.e.$\big({{P^{I,A}_{K_{[1,2]}}}(t)}\big)$), along with their associated transmission rates ($\big({\beta^{I,A}_{K_{[1,2]}}(t)}\big)$), on the corresponding multiplex layers. From the figure, it can be seen, that the probabilities and the respective transmission rates for both the processes (i.e. disease and awareness) have dependence upon the intra-layer node degree. The comparison between the degree based progression characteristics of the two processes on the infection (${L_{1}}$) and awareness layers (${L_{2}}$) highlights, the difference in the parametric dependence of marginal probabilities upon the respective transmission rates. For the infection layer (${L_{1}}$), the increase in the intra-layer node degree (${{K_{1}}:{min\rightarrow max}}$) increases the infection probability (${{P^{I}_{K_{1}}}:{0\rightarrow 1}}$), but it lowers the rate of transmission (${{\beta}^{I}_{K_{1}}:1\rightarrow 0}$) in each of the aforementioned domains. By contrast, the increase in the intra-layer node degree (${K_{2}: min \rightarrow max }$) on the awareness layer (${L_{2}}$), results in the simultaneous increase in the awareness probability (${P^{A}_{K_{2}}:0\rightarrow 1}$) and the respective rates of transmission (${ \beta^{A}_{K_{2}}:0\rightarrow 1}$) for each of the three domains.

  From these results, it can be inferred that, the awareness resulting from disease information and intra-layer node degree have contrasting roles in the spread of disease. The spread of awareness (i.e.${P^{A}_{K_{2}}:0\rightarrow 1}$) tend to reduce the infection probability (i.e.${{P^{I}_{K_{1}}}:{1\rightarrow 0}}$) by lowering the respective disease transmission rates (${\beta^{I}_{K_{1}}: 1\rightarrow 0 }$), but an increase in the intra-layer node degree (i.e. ${K_{2}:min \rightarrow max }$ ) tends to enhance the infection probability (i.e.${{P^{I}_{K_{1}}};{0\rightarrow 1}}$). Thus the effective infection probability of node is a result of these two mutually competing factors. For lower degree nodes, the awareness predominates over the effect of intra-layer node degree, and as a result, the infection probability of these nodes is lower even with a high transmission rate. In contrast, for nodes with a high intra-layer degree, the awareness cannot reduce the infection probability but can only lower the corresponding transmission rates. The cumulative effect of multiple simultaneous interactions due to high degree results in high infection probability (i.e.${{P^{I}_{K_{1}}}{\approx 1}}$). However, the high infection probability also implies greater self-awareness, which in turn increases the rate of transmission of disease information (${\beta^{A}_{K_{2}}: 0\rightarrow 1}$) and facilitates its spread on the awareness layer (i.e.${{P^{A}_{K_{2}}}{\approx 1}}$). Thus it is evident, that the intra-layer linking pattern of nodes has a significant effect on the temporal evolution of these processes over the respective layers of the multiplex.

  As previously stated, the layers of multiplex network can have similar or distinct internal structures consequently the degree of nodes in each layer can be same or different. Thus, it is worthwhile to examine the effect of both relative and absolute degree of the participating nodes on the progression characteristics of these processes. Fig-7 highlights the effect of relative difference in the inter-layer node degree (i.e. ${K_{1}-K_{2}}$) on the temporal evolution of the infection and awareness marginal probabilities along with the respective rates of transmission for each of the three aforementioned cases. The variable ${\delta}$ (=${(K_{1}-K_{2})}$) has been used to describe the relative difference in the inter-layer degree of nodes. The temporal evolution of infection and awareness probabilities together with the respective rates for the previously outlined three cases can be further segregated in terms of this variable ($\delta$) into following sub-categories: The positive value of the variable $\delta$, (i.e. ${(K_{1}-K_{2})>0}$) corresponds to the first sub-category for which the relative degree of nodes in the infection layer (${ L_{1}}$) is greater than the awareness layer (${L_{2}}$). Similarly, the near zero value of the variable $\delta$, (i.e.${(K_{1}-K_{2}) \approxeq 0}$) corresponds to the second sub-category for which the relative degree of nodes is equal in either of the layers. Lastly, the sub-zero value of variable $\delta$ (i.e.${(K_{1}-K_{2})<0}$), corresponds to the third category for which the relative degree of nodes in the awareness layer (${L_{2}}$) is greater than that in the disease layer (${L_{1}}$).

 From the figure, it is evident, that the infection and awareness probabilities of the nodes along with their transmission rates have a dependence on the relative degree of the nodes. The analysis of the disease progression on the infection layer (${L_{1}}$) shows that the finite awareness lowers the disease transmission rates on layer ${\beta^{I}_{\delta}}$. Further, the extent of this decrease has an inverse dependence on the modulus of relative difference in the node degrees (i.e. $\mid \delta \mid$). From the figure it can also be seen that the infection probability of the nodes is higher for the sub-category of nodes belonging to $\delta \geq 0$ than for $\delta \leq 0$ sub-category.  Moreover for ($\delta \geq 0$), the infection probabilities of the corresponding nodes increases with an increase in the relative difference in the degree of the nodes.  However for ($\delta \leq 0$) corresponding to cases ({\bf 2}) and ({\bf 3}), an increase in the magnitude of relative difference of the node degree (i.e. $-\delta$), initially decreases the infection probability but any further increase results in variable levels of infection probabilities. Thus, the effect on infection probability is not symmetric about the relative difference in the inter-layer degree of nodes.

     From an examination of the dynamics on the awareness layer (${L_{2}}$), it can be seen that for $\delta \geq 0$, which corresponds to larger relative degree of nodes on the disease layer ($K_{1} > K_{2}$), the higher infection probabilities results in greater self awareness. The larger self-awareness increases the corresponding rate of transmission (i.e. $\beta^{A}_{\delta}\approx 1$) and results in high awareness probabilities (i.e. ${P^{A}_{K_{2}}\approx 1}$). The high awareness probability in turn lowers the respective rates of disease transmission ($\beta^{I}_{\delta}$). However for $\delta \leq 0$ (i.e.${K_{1} < K_{2}}$), the relatively higher degree of the nodes in the awareness layer in comparison to the infection layer facilitates the spread of awareness. The resultant larger awareness probability in turn simultaneously reduces the disease transmission rates ($\beta^{I}_{K_{1}}$) and the infection probability. Therefore, for these nodes the magnitude of awareness transmission rates (${\beta^{A}_{K_{2}}}$) is relatively greater than the disease transmission rates (${\beta^{I}_{K_{1}}}$). Thus from this analysis, it can be concluded that the relative difference in the inter-layer degree ($\delta$) also has a significant effect on the infection and awareness probabilities as well on the modulation intensity of respective rates.

    The observed asymmetry in the infection probability with relative difference in the degree of the nodes (${P^{I}_{\delta}}$) for the case $\delta \leq 0$, in  Fig-3 can be understood by further analyzing its simultaneous dependence on the absolute degree of the nodes (i.e. ${K_{1},K_{2}}$). Fig-8, describes the dependence of the infection and awareness probabilities on the absolute degree of nodes (i.e. $ P^{I}(K_{1},K_{2})$ and $P^{A}(K_{1},K_{2})$). From Fig-8, it can be inferred that for the cases ({\bf 2}) and ({\bf 3}) in the parametric regime corresponding to $\delta \leq 0 $ in Fig-3, the variation in the infection probability depends not just on the relative difference in the degree of nodes but has additional dependence on the absolute degree of nodes. This inference implies that the absolute number of physical contacts in the disease layer has a dominant effect on the infection probability of the nodes where the high degree of awareness can only slow down the spread of disease but cannot subdue it. Thus, from these results it can be concluded that the new framework provides a deeper insight into the role of network structure modulated self-consistent temporal variation of transmission rates on the concurrent spread of disease coupled awareness over a multiplex networks.

\section{Summary}
To summarize, a novel framework has been proposed to elucidate the effect of inductive coupling on the co-propagation of disease and its associated awareness over a two layered multiplex network. The spread of disease due to the persistent physical contacts over the infection layer is considered to be concurrent with the spread of its corresponding disease information induced awareness by the virtual contacts over the awareness layer. The dynamics of these mutually competing processes results in the self-consistent temporal variation of the respective rates of transmission for each of these processes. The modulation intensity of the respective transmission rates has an implicit dependence over the intra-layer and inter-layer linking pattern of nodes. Such a temporal variation in the transmission rates can affect the critical properties for the spread of disease and its awareness. From the extensive numerical and theoretical work it has been shown that the progression characteristics of these processes depend upon the choice of minimum rates of transmission for each layer. For the case, when the minimum value of the transmission rates for each of the layer is chosen to be less than its critical value, the system doesn't exhibit an equilibrium state. But it is rather characterized by alternating repetitive cycles of high level of infection followed by high awareness. However, the choice of minimum value of the transmission rates to be greater than or equal to the critical value results in equilibrium states of the  system. In the equilibrium state, there is a simultaneous prevalence of disease and its awareness. From the  results, it has been further found that the absolute infection probability of the nodes is predominantly dependent upon the degree of the nodes in the disease layer. The findings of the present work can account for the real life scenario wherein an individual can have high infection probability even with high level of disease awareness and low rate of disease transmission owing to its larger connectivity. Moreover, it is also shown that in the continuous time limit, the co-progression of these interacting processes can be described in terms of coupled damped and driven oscillator equations. The self-consistent driving force is shown to depend upon the internal structure of the multiplex networks.

\acknowledgments
The authors V.S. and Y.Z. Acknowledge support from the National Nature Science Foundation Committee (NSFC) of China under Project No. 61573119 and a Fundamental Research Project of Shenzhen under Project Nos. JCYJ20140417172417109, and JCYJ20140417172417090. A.S. thanks the Indian National Science Academy (INSA) for their support under the INSA Senior Scientist Fellowship scheme.


\newpage
\begin{center}
{\bf FIGURE CAPTIONS}
\end{center}
\noindent
Figure(1): The figure illustrates the inductively coupled co-evolution of the infection and its corresponding awareness over the $L_{1}$ and $L_{2}$ layers of the multiplex networks. The individuals simultaneously interacting in different modes over the respective layers of the multiplex network are represented in terms of the same coloured nodes. On the infection layer ($L_{1}$), $\beta^{I}_{i}(P^{A}_{i}(t))$ is the awareness probability dependent rate of transmission of infection of the $i^{th}$ node. Similarly, on the awareness layer ($L_{2}$), $\beta^{A}_{i}(P^{I}_{i}(t))$ is the infection probability dependent rate of transmission of awareness of the $i^{th}$ node.        

\noindent
Figure(2):Probability tree for the transitions of states on infection and awareness layers of the multiplex.

\noindent
Figure(3):Description of the temporal evolution of un modulated infection and awareness probabilities  of the nodes as a function of corresponding rates of transmission.

\noindent
 Figure(4):Description of the temporal evolution of infection and awareness probabilities  of the nodes as a function of corresponding rates of transmission.

\noindent
Figure(5):Phase Diagram describing the temporal evolution of averaged infection $\rho_{avg}^I$ and awareness $\rho_{avg}^A$ probabilities of the nodes as a function of corresponding rates of transmission($\beta_{avg}^{I,A}$) for three different cases.

\noindent
Figure(6): Temporal evolution of infection and awareness probabilities  as function of node degree$({\bf K_{1},K_{2}})$ and respective rates of transmission($\beta^{I,A}_{K_{1},K_{2}}$).

\noindent
Figure(7): Temporal evolution of the infection and awareness probabilities as a function of the relative inter-layer node degree($\delta(K_{1}-K_{2})$).

\noindent
Figure(8):Temporal evolution of the infection and awareness probabilities as a function of the absolute node degree on each of the multiplex layers($(K_{1},K_{2})$).

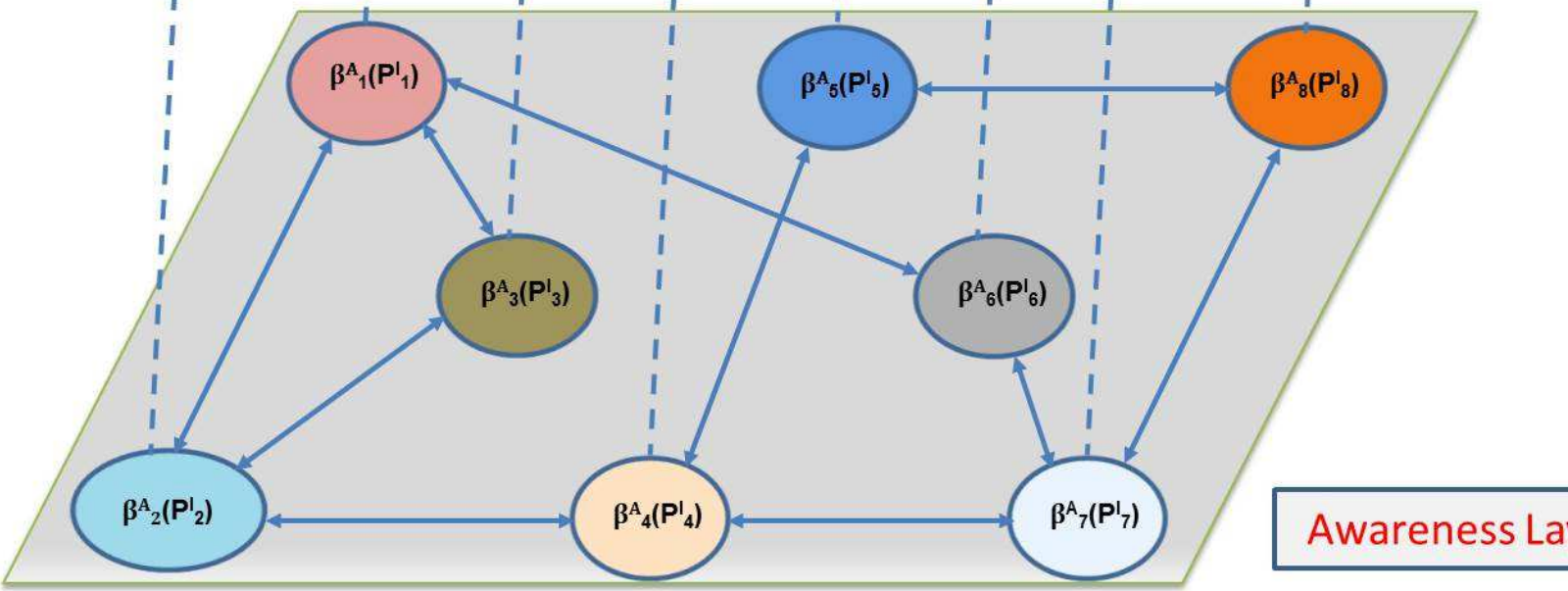
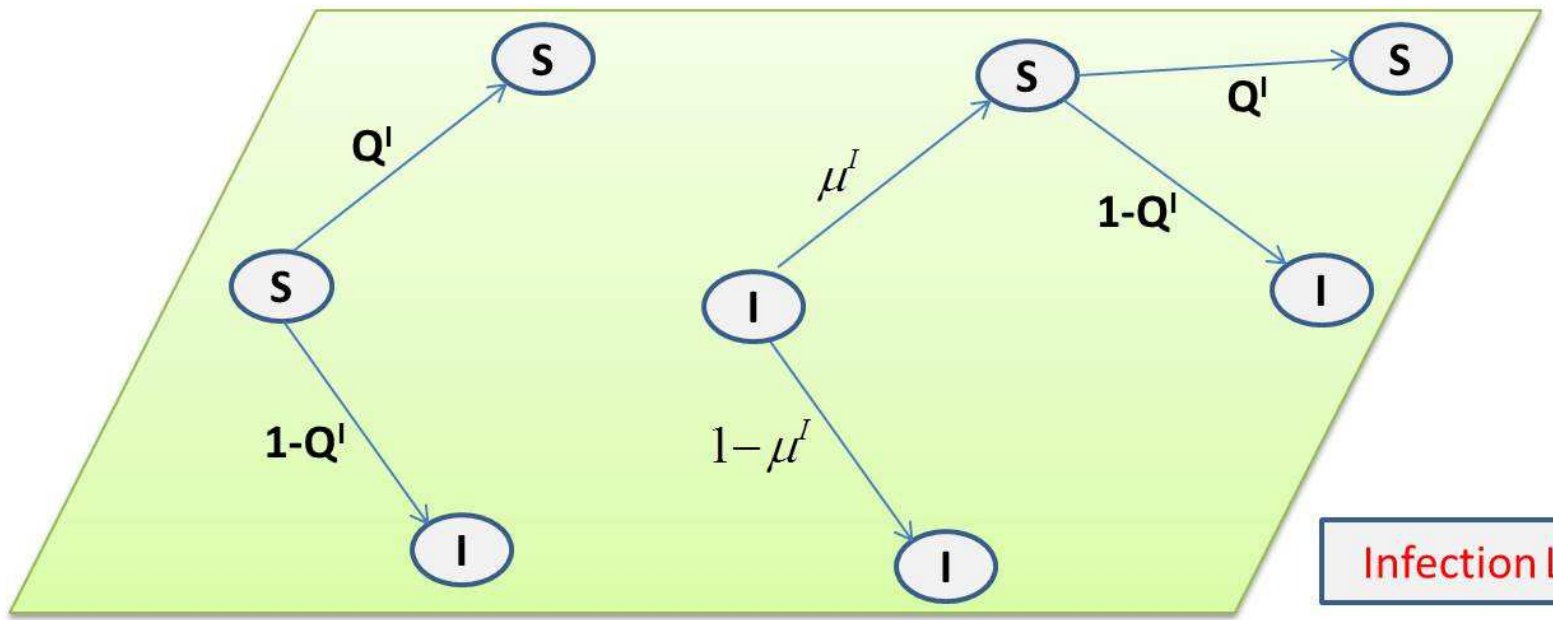
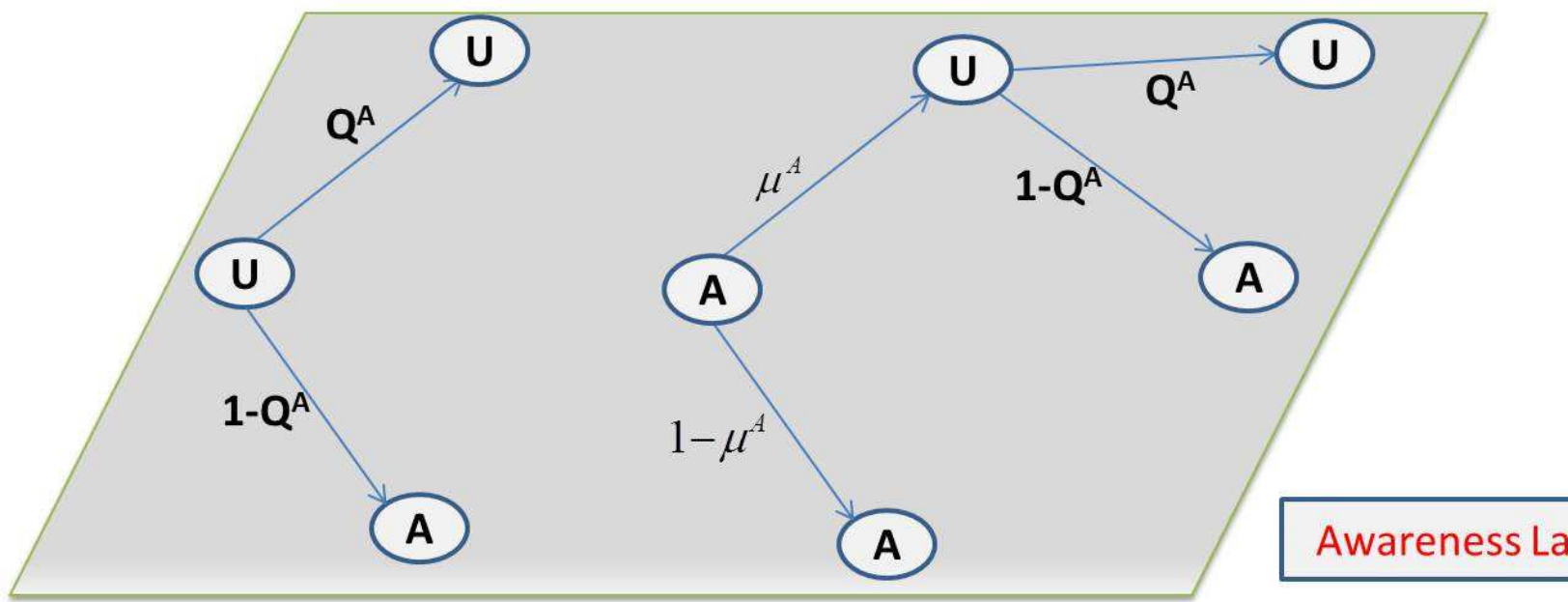
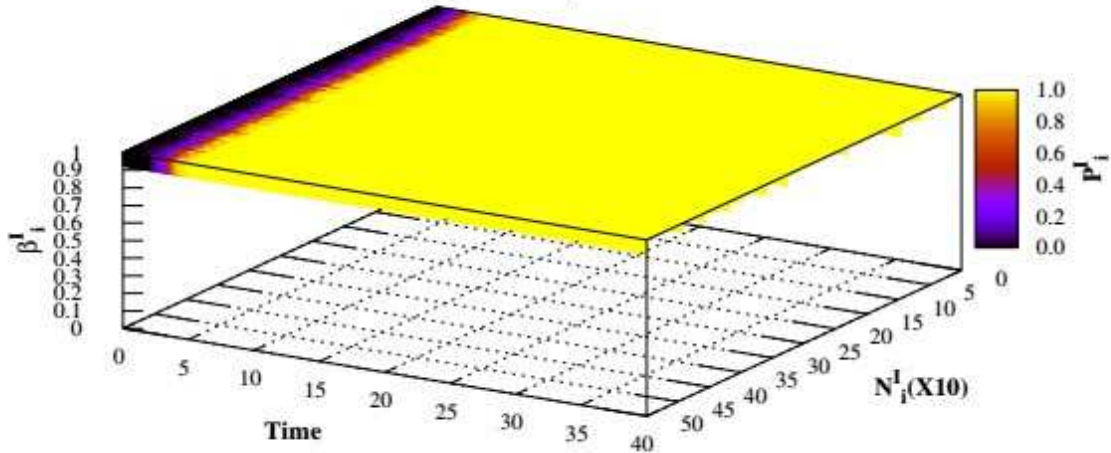
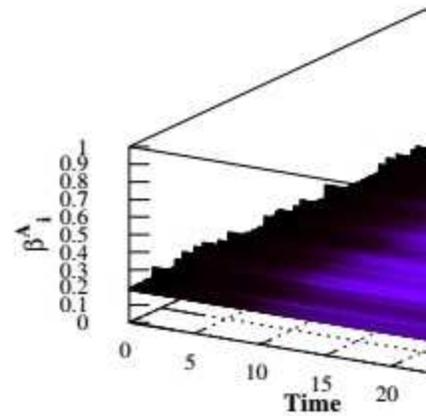
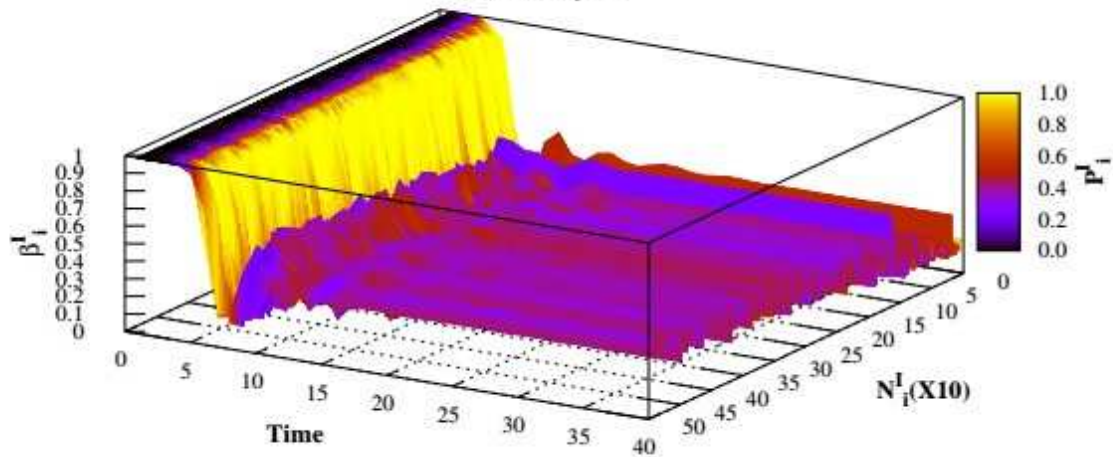
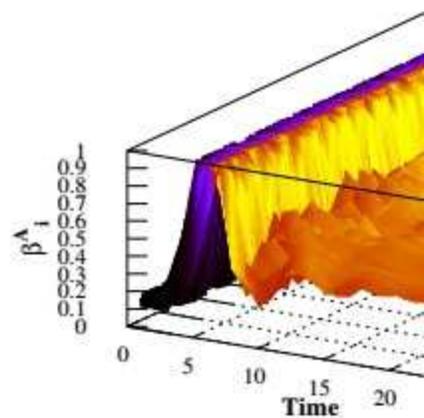
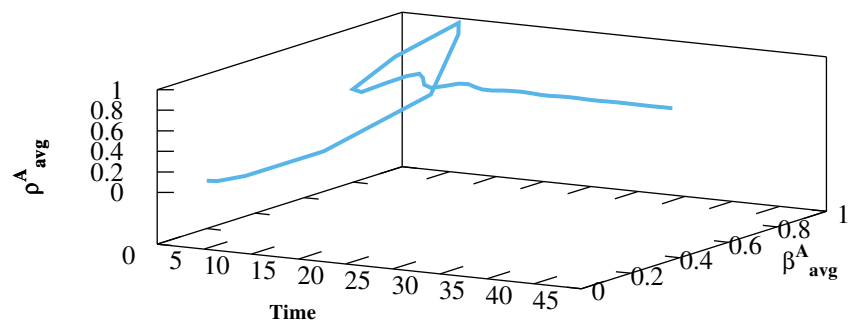
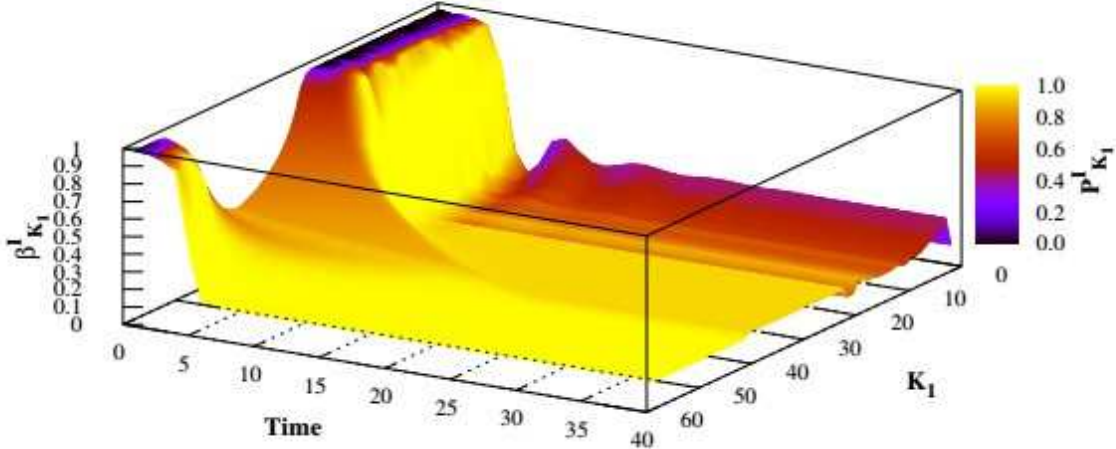
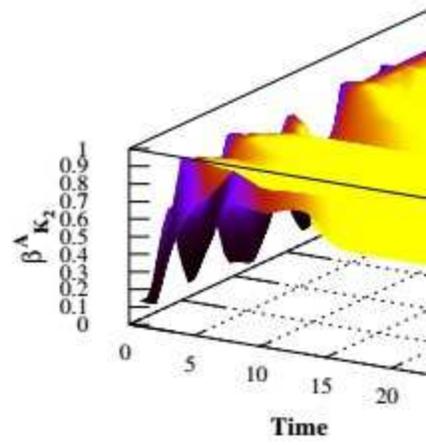
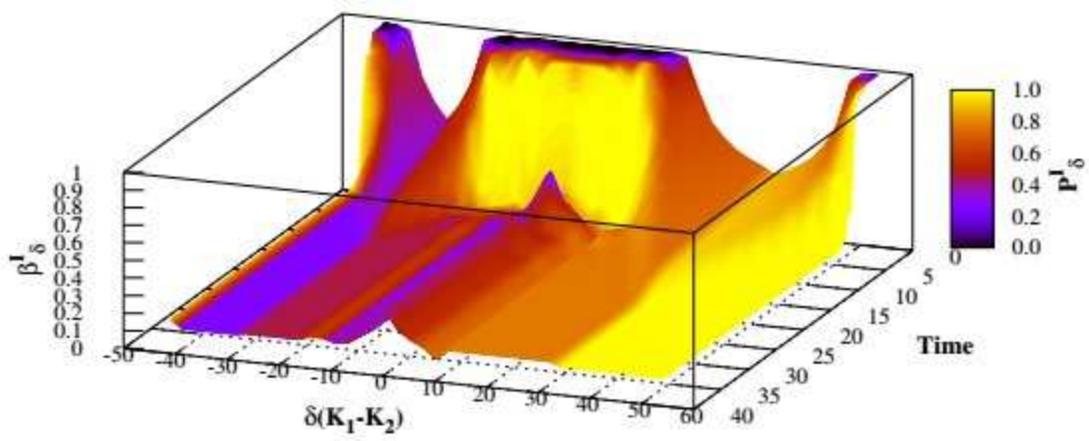
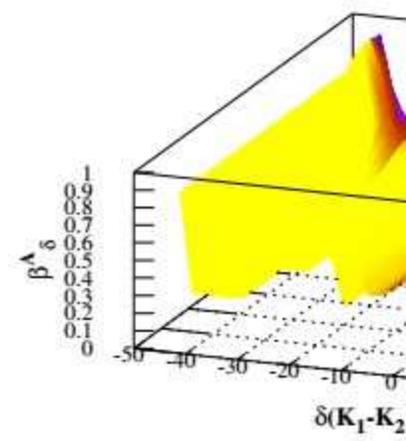
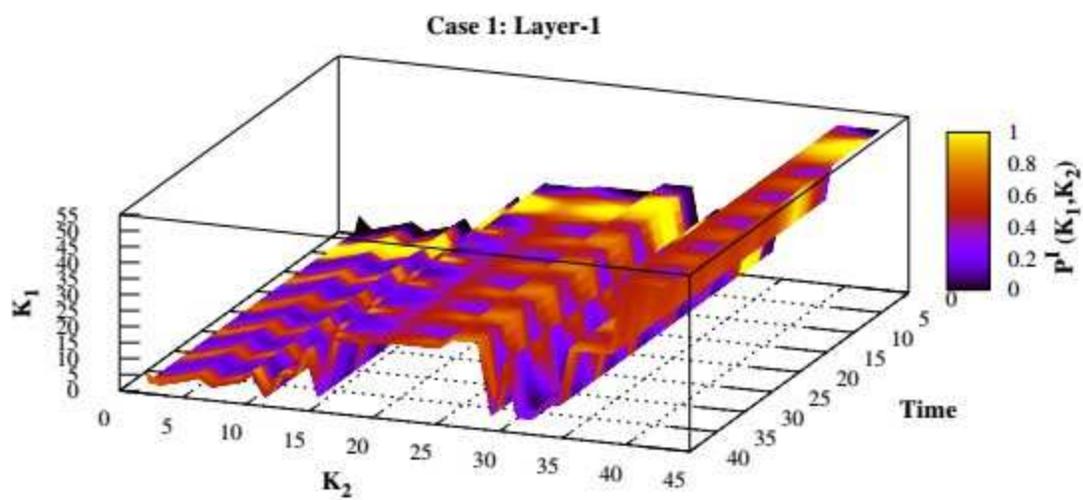
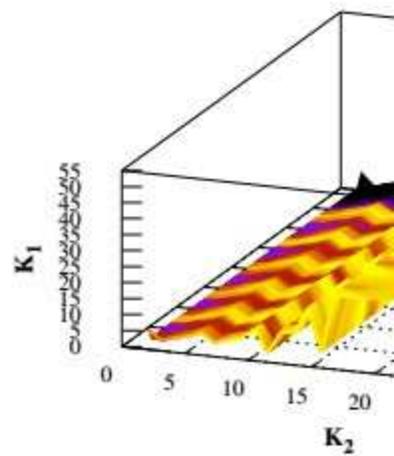
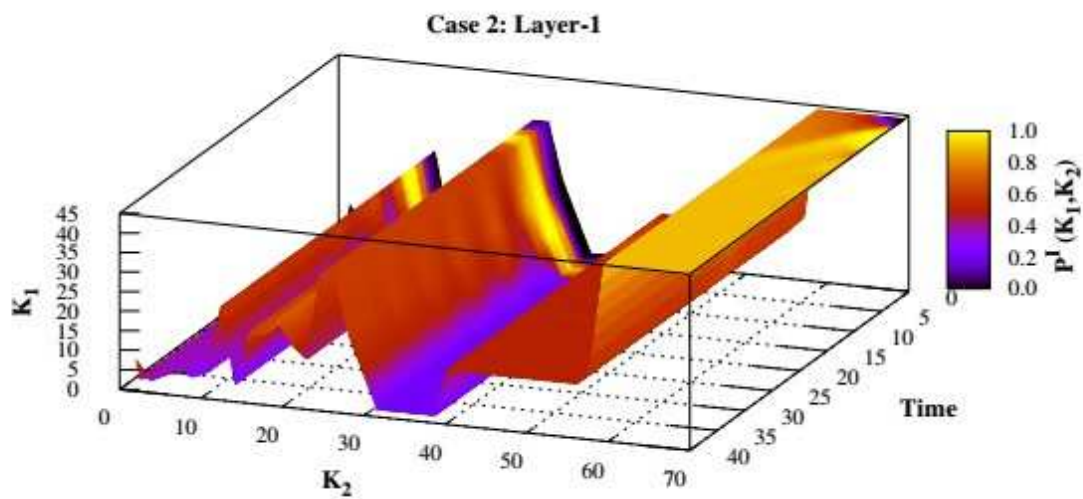
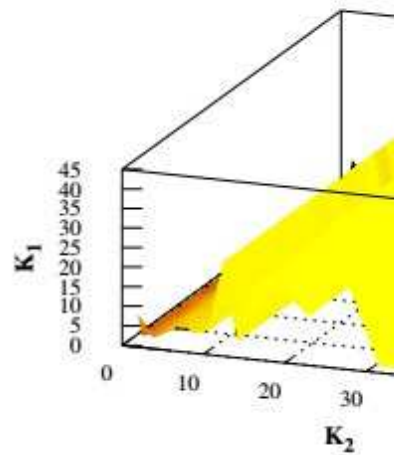
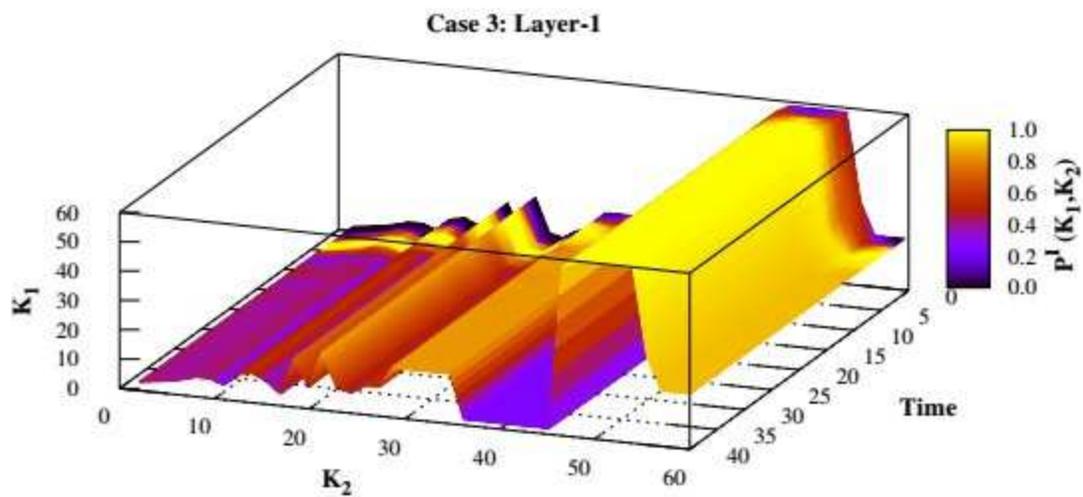
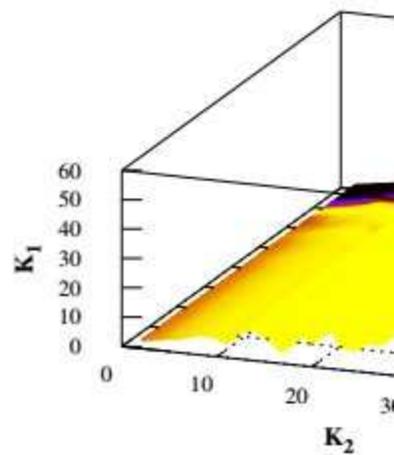
\end{document}